\documentclass[prl,showpacs,twocolumn,aps]{revtex4}%
\usepackage{graphicx}
\usepackage{epsfig}
\usepackage[english]{babel}
\usepackage{amsmath}
\usepackage{amsfonts}
\usepackage{amssymb}%
\begin{document}
\title{Non-KPZ modes in two-species driven diffusive systems}
\author{V. Popkov, J. Schmidt}
\affiliation{Institut f\"{u}r Theoretische Physik, Universit\"{a}t zu K\"{o}ln, Z\"ulpicher Str. 77,
50937 Cologne, Germany.}
\author{G.M.~Sch\"utz}
\affiliation{Institute of Complex Systems II, Theoretical Soft Matter and Biophysics,
Forschungszentrum J\"ulich, 52425 J\"ulich, Germany}
\affiliation{Interdisziplin\"ares Zentrum f\"ur Komplexe Systeme, Universit\"at
Bonn, Br\"uhler Str. 7, 53119 Bonn, Germany}

\begin{abstract}
Using mode coupling theory and dynamical Monte-Carlo simulations we investigate the
scaling behaviour of the dynamical structure function of a two-species
asymmetric simple exclusion process, consisting of two coupled single-lane
asymmetric simple exclusion processes. We demonstrate the appearence of a superdiffusive mode 
with dynamical exponent $z=5/3$ in the density fluctuations, along with a KPZ mode with 
$z=3/2$ and argue that this phenomenon is generic for short-ranged 
driven diffusive systems with more than one conserved density. When the dynamics 
is symmetric under the interchange of the two lanes
a diffusive mode with $z=2$ appears instead of the non-KPZ superdiffusive mode.

\end{abstract}
\date{\today }

\pacs{05.60.Cd, 05.20.Jj, 05.70.Ln, 47.10.-g}
\maketitle


Transport in one dimension has been known for
a long time to be usually anomalous \cite{Alde67,Erns76}. Signatures of this behaviour are a
superdiffusive dynamical structure function and a power law divergence of transport 
coefficients with system size, characterized by universal critical exponents.  Unfortunately,
however, despite a vast body of work, analytical results for model systems have remained scarce 
and numerical results are often inconclusive.  Therefore 
the exact calculation \cite{Prae02} of the dynamic structure function for the universality class of the 
Kardar-Parisi-Zhang-equation \cite{Kard86} with dynamical exponent $z=3/2$ some ten years ago 
came as a major breakthrough. This function was obtained for a specific driven diffusive system,
the asymmetric simple exclusion process which has a single conserved density
and hence a single mode, the KPZ-mode.
By virtue of universality many results for other types of systems such as 
growth models \cite{Prae04}, hard-core particle systems \cite{Gras02}
or anharmonic chains \cite{Delf07,Bern13} can thus
be understood in terms of the KPZ universality class. 

More recently it was established in the context of anharmonic chains \cite{Spoh13} 
and very general short-ranged one-dimensional Hamiltonian systems \cite{vanB12} 
that in the presence of more than one conserved quantity the dynamics is richer 
and other modes have to be expected. In particular, in systems with three conservation laws 
a heat mode with $z=5/3$ (corresponding to a divergent finite-size heat conductivity 
$C\propto L^{1/3}$) or a diffusive mode with $z=2$ may be present besides two KPZ 
modes. The main assumption underlying these conclusions is that the relevant slow 
modes are given by the long wave length relaxation behaviour of the conserved
quantities \cite{Erns76,vanB12}.

Going back to driven diffusive systems we note that somewhat surprisingly, there is little information 
about the dynamical structure function in driven diffusive systems with more than one conservation law. 
In one dimension these systems are known to exhibit extremely 
rich stationary and dynamical behaviour and they serve widely as
paradigmatic models for the detailed study of non-equilibrium phenomena.
In view of this it is of interest
to explore the transport properties of such systems, in particular which modes govern the fluctuations
of the locally conserved slow modes.

In this spirit Ferrari et al. \cite{Ferr13} studied very recently a two-species exclusion process, 
using mode-coupling theory and Monte-Carlo simulations, and found two very clean KPZ-modes,
but no other modes. For a similar model, exact finite size scaling analysis of the spectrum indicates
a dynamical exponent $z=3/2$ \cite{Arit09}. In older work on other lattice gas models 
with two conservation laws the presence of a 
KPZ mode and a diffusive mode was observed \cite{Das01,Rako04}. So far there has been no
indication of the existence of a heat mode. In the light of the work \cite{vanB12,Spoh13} on
short-ranged Hamiltonian systems this is intriguing and
raises the question whether a mode with $z=5/3$ can exist in driven diffusive systems, and, if yes, 
how many conservation laws are required to generate it.
In this letter we 
answer these questions by using the mode coupling theory developed 
in \cite{Spoh13,Ferr13} for
non-linear fluctuating hydrodynamics and by confirming the analytical findings with
Monte-Carlo simulations of a two-species asymmetric simple exclusion process.
It will transpire that
a superdiffusive $z=5/3$ mode along with a KPZ mode exists and that two conservation laws are sufficient to 
generate the phenomenon. Also a KPZ mode along with a diffusive mode can occur
on a line of higher symmetry, which is reminiscent of a similar behaviour in anharmoic
chains \cite{Spoh13}.

\begin{figure}[ptb]
\begin{center}
\includegraphics[
scale=0.45
]{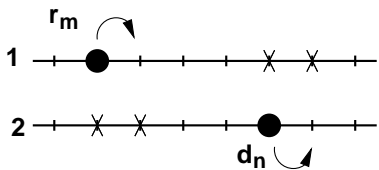}
\end{center}
\caption{Schematic representation of the two-lane totally asymmetric simple exclusion  process. A particle on lane 1 (2)
hops to the neighbouring site (provided this target site is empty) with rate $r_m$ ($d_n$), where $m$ ($n$) is the number
of particles on the adjacent sites of the other lane that are marked by a cross.}%
\label{Figmodel}%
\end{figure}

We consider the following stochastic lattice gas model. Particles hop randomly
on two parallel chains with $N$ sites each, without exchanging the lane, unidirectionally and
with a hard core exclusion and periodic boundary conditions. 
We denote the particle occupation number on site $k$ in
the first (upper) lane by $n_{k}$ , and on the other lane by $m_{k}$. A
hopping event from site $k$ to site $k+1$ on the same lane may happen if site
$k$ is occupied and site $k+1$ is empty. The rate of hopping
depends on the sum of particle numbers at sites $k,k+1$ on the adjacent lane 
as follows (Fig.~\ref{Figmodel}):
Let us denote the sum of particles on the sites $k,k+1$ at the
second lane as $m:=m_{k}+m_{k+1}$. 
Then the rates $r_{m}$ of hopping from site $k$
to site $k+1$ on the first lane are given by
\begin{equation}
r_{m}=1+\frac{\gamma m}{2}, \label{r_m}%
\end{equation}
where $\gamma$ is a coupling parameter. Analogously, the
rates $d_{n}$ of hopping from site $k$ to site $k+1$ on the
second (lower) lane are given by
\begin{equation}
d_{n}=b+\frac{\gamma n}{2} \label{d_n}%
\end{equation}
with the sum of
particles $n:=n_{k}+n_{k+1}$ on the sites $k,k+1$ at the first lane. 
Since there is only hopping within lanes, the
total number of particles $M_i$ in each lane is conserved. The parameter $b$ 
makes the lanes inequivalent.
Since the rates have to be nonnegative, if follows that $\gamma
\geq-\min(1,b)$. For $b=1$ we recover the two-lane model of \cite{Popk03}.

The model in a more general multilane geometry was
introduced in \cite{Popk04}.
It was shown that the choice of rates (\ref{r_m}), (\ref{d_n})
results in a stationary distribution which is a product measure, both between lanes
and between the sites. For the two-lane system that 
we study here this leads to stationary currents
\begin{eqnarray}
j_{1}(\rho_{1},\rho_{2})  &  = & \rho_{1}(1-\rho_{1})(1+\gamma\rho_{2})\nonumber \\
\label{j1j2} 
j_{2}(\rho_{1},\rho_{2})  &  = & \rho_{2}(1-\rho_{2})(b+\gamma\rho_{1}) 
\end{eqnarray}
where $\rho_{1,2}=M_{1,2}/N$ are the densities of particles in the first and
second lane respectively.
Notice that a product measure corresponds to a grandcanonical ensemble
with a fluctuating particle number. These fluctuations are described by the
symmetric compressibility matrix $C$ with matrix elements 
\begin{equation}
\label{C}
C_{ij} = \frac{1}{N}
<(M_i - \rho_i N) (M_j - \rho_j N)> = \rho_i(1-\rho_i)\delta_{i,j} .
\end{equation}

The starting point for investigating the large-scale dynamics of this microscopic
model is the
system of conservation laws
$\partial_t \rho_i(x,t) + \partial_x j_i(x,t) =0$
where $\rho_i(x,t)$ is the coarse-grained local density of component $i$, and $j_i(x,t)$ is the
associated current given as a function of the local densities by (\ref{j1j2}) \cite{Kipn99}.
These equations can be written in vector form as
\begin{equation}
\label{hyper}
\frac{\partial}{\partial t} \vec{\rho} + A \frac{\partial}{\partial x} \vec{\rho} = 0
\end{equation}
where $\vec{\rho}$ is the column vector with the densities as entries and $A$ is the
current Jacobian with matrix elements $A_{ij} = \partial j_i / \partial \rho_j$.
Its eigenvalues $c_i$ are the characteristic velocities which on microscopic scale
are the speeds of local perturbations \cite{Popk03}. The matrix $S=AC$ is
symmetric \cite{Gris11} 
which guarantees that the system (\ref{hyper}) is hyperbolic \cite{Toth03}.

Eq. (\ref{hyper}) describes the deterministic time evolution of the density under Eulerian scaling.
The effect of fluctuations, which occur on finer space-time scales, can be 
captured by adding phenomenological white noise terms $\xi_i$
and taking the non-linear
fluctuating hydrodynamics approach together with a mode-coupling analysis of the non-linear
equation \cite{Ferr13}. In this framework
one expands the local densities around their long-time stationary values 
$\rho_i(x,t) = \rho_i + u_i(x,t) $ and transforms to normal modes $\vec{\phi} = R \vec{u}$
where $A$ is diagonal. The transformation matrix $R$ uniquely defined by
$RAR^{-1} = \mathrm{diag}(c_i)$ and the normalization condition $RCR^T = 1$.
Keeping terms to first non-linear order
yields
\begin{equation}
\partial_t \phi_i =  - \partial_x \left( c_i \phi_i + \frac{1}{2} \langle  \vec{\phi}, G^{(i)} \vec{\phi}  \rangle - \partial_x  (D \vec{\phi})_i
+ (B \vec{\xi})_i \right).
\end{equation}
Here the angular brackets denote the inner product in component space and 
\begin{equation}
G^{(i)} =  \frac{1}{2} \sum_j R_{ij} (R^{-1})^T H^{(j)} R^{-1}.
\end{equation}
are the mode coupling matrices obtained from the Hessian
$H^{(i)}$ with matrix elements $\partial^2 j_i /\partial \rho_j \partial \rho_k$.
The matrices $D$ (transformed diffusion matrix) and $B$ (transformed noise strength) 
do not appear explicitly below.

For strictly hyperbolic systems the normal modes have different speeds and hence their interaction
becomes very weak for long times. Thus, by identifying $\phi_i$ with the gradient of a height variable 
one has to leading order generically two decoupled KPZ-equations
with nonlinearity coefficients $G^{(i)}_{ii}$. The other diagonal
elements $G^{(i)}_{jj}$ provide the leading corrections to the KPZ modes, the offdiagonal 
elements result in subleading corrections.
We point out the scenarios relevant to our model, as predicted by mode-coupling theory.
(i) If both $G^{(1)}_{11}$ and $G^{(2)}_{22}$ are non-zero we expect two KPZ modes with $z=3/2$. (ii) On the other hand, if e.g. 
$G^{(1)}_{11}=0$, but $G^{(1)}_{22}\neq 0$ and $G^{(2)}_{22}\neq 0$, then mode coupling theory predicts
mode 1 to be a superdiffusive mode with $z=5/3$ and mode 2 to be KPZ. (iii) Finally, if both $G^{(1)}_{11}=G^{(1)}_{22}=0$ but
$G^{(2)}_{22}\neq 0$
then mode 1 becomes diffusive, while mode 2 is KPZ. 

For our system, the explicit forms of $A$ and $H^{(i)}$ are%
\begin{equation}
A=%
\begin{pmatrix}
(1+\gamma\rho_{2})(1-2\rho_{1}) & \gamma\rho_{1}(1-\rho_{1})\\
\gamma\rho_{2}(1-\rho_{2}) & (b+\gamma\rho_{1})(1-2\rho_{2})
\end{pmatrix}
\label{A}%
\end{equation}%
\begin{equation}
H^{(1)}=%
\begin{pmatrix}
-2(1+\gamma\rho_{2}) & \gamma(1-2\rho_{1})\\
\gamma(1-2\rho_{1}) & 0
\end{pmatrix}
\end{equation}%
\begin{equation}
H^{(2)}=%
\begin{pmatrix}
0 & \gamma(1-2\rho_{2})\\
\gamma(1-2\rho_{2}) & -2(b+\gamma\rho_{1})
\end{pmatrix}
\end{equation}
To prove that all three scenarios (i) - (iii) can be realized, we choose $\rho_1=\rho_{2}=:\rho$ and for
convenience we set $\gamma=1$.

Consider first $b=2$. Then 
\begin{equation}
R=R_{0}\left(
\begin{array}
[c]{cc}%
1-\rho & -\rho\\
\rho & 1-\rho
\end{array}
\right)
\end{equation}
where $R_{0}^{-1}=\sqrt{\rho(1-\rho)(\rho^2+(1-\rho)^2)}$.
The characteristic velocities are
\begin{equation}
c_{1}=1-\rho-3\rho^{2}, \quad c_{2}=2-3\rho-\rho^{2} \label{c1c2}%
\end{equation}

The matrices
$G^{(1)},G^{(2)}$ are symmetric and have matrix elements
$G^{(1)}_{11} = -2 g_{0} (  6\rho^{4}-8\rho^{3}+5\rho^{2}+\rho-1)$,
$G^{(1)}_{12} = G^{(1)}_{21} = g_{0} (4\rho^{3}-10\rho^{2}+8\rho-1)$, 
$G^{(1)}_{22} = - 2 g_0 \rho(1-\rho)(  2\rho^{2}-6\rho+3)$
and
$G^{(2)}_{11} = 4 g_{0} \rho(1-\rho)$, 
$G^{(2)}_{12} = G^{(2)}_{21} = - g_{0} (1-2\rho^{2})^{2}$, 
$G^{(2)}_{22} = 4 g_0 (1 - 3 \rho(1-\rho))$
with
$g_{0} = - 1/2 \left[ \rho(1-\rho) / (1-  2\rho(1-\rho))^3 \right]^{1/2} $.
Therefore, generically condition (i) for the presence of two KPZ 
modes is satisfied. 
However,  while $G^{(2)}_{11} \neq 0$  and $G^{(2)}_{22} \neq 0$
$\forall \rho \in (0,1)$,
the self coupling coefficient $G^{(1)}_{11}$ changes sign at
$\rho^{\ast} =0.45721\dots$. Since $G^{(1)}_{22}(\rho^\ast) \neq 0$,
the condition for case (ii), KPZ mode plus superdiffusive non-KPZ mode, is thus satisfied at density $\rho=\rho^\ast$.
In fact, diagonalizing $A$ for arbitrary densities $\rho_1,\rho_2$ one can show
that for $b\neq1$ there is a curve in the space of densities where condition (ii) is
satisfied. On the other hand,
there is no density where condition (iii), $G^{(1)}_{11}=G^{(1)}_{22}=0$, is satisfied.
Indeed, numerical inspection of the mode coupling matrices for several
other parameter choices of $\gamma$ and $b$ suggests that condition (iii)
cannot be satisfied when $b\neq 1$.

Next we study $b=1$. In this case the system is symmetric under interchange of the
two lanes, which is reflected in the relation $j_2(\rho_1,\rho_2)=j_1(\rho_2,\rho_1)$
for the currents (\ref{j1j2}). Calculating the mode coupling matrices for $\rho_1=\rho_{2}=:\rho$ 
and $\gamma=1$ yields
\begin{equation}
G^{(1)}= \tilde{g}_0 (1+\rho) \left(
\begin{array}
[c]{cc}%
0 & 1 \\
1 & 0
\end{array}
\right), 
G^{(2)} = \tilde{g}_0 \left(
\begin{array}
[c]{cc}%
2-\rho & 0\\
0 & 3\rho
\end{array}
\right)
\end{equation}
with $\tilde{g}_0 = - \sqrt{2\rho(1-\rho)}$.
Interestingly, in this case condition (iii) is satisfied for all $\rho$,  i.e., mode 1 is expected to be 
diffusive and mode 2 is KPZ. The occurrence of a diffusive mode is somewhat counter-intuitive as
both particle species interact and hop in a totally asymmetric fashion which in the case
of the AHR-model prevents the existence of a diffusive mode \cite{Ferr13}.

In order to check the predictions of mode coupling theory we performed
dynamical Monte-Carlo simulations, using a random sequential update
where in each step a random site is chosen uniformly and jumps are performed
with probabilities determined by normalizing the jump rates (\ref{r_m}),
(\ref{d_n}) by the largest jump rate, provided the target site is empty. 
A full Monte Carlo time step then corresponds
to $2N$ such update steps. The initial distribution is sampled from the uniform
distribution, except for the occupation number at site $N/2$ which is determined
according to the normal modes given by the transformation matrix $R$.
Averages are performed over up to $10^8$ realizations of the process and $N=200 \dots 300$.
In order to measure the dynamical exponent, we compute the first and second
moment of the dynamical structure function, from which we obtain the variance 
$\sigma(t)  = <X^2(t)> -<X(t)>^2\propto t^{2/z}$
of the density distribution as a function of time.

In order to test the existence of a superdiffusive non-KPZ mode we have chosen
$\gamma=-0.52588$ and $b=1.3$. This yields $G_{22}^{(2)}=0$ at $\rho_{1}^{\ast}
=\rho_{2}^{\ast}=0.5500003 \approx 55/100$. 
The matrices $G^{(1)},G^{(2)}$ become
\[
G^{(1)}=\left(
\begin{array}
[c]{cc}%
0.2950 & 0.0717\\
0.0717 & 0.3157
\end{array}
\right), \quad
G^{(2)}=\left(
\begin{array}
[c]{cc}%
0.0706 & 0.2972\\
0.2972 & 0
\end{array}
\right)
\]
which means that mode 2 is expected to be a non-KPZ mode and mode 1 is KPZ.
The corresponding characteristic
velocities are $c_{1}(\rho^{\ast})=-0.2171$, $c_{2}(\rho^{\ast})=0.0449$.
and the eigenvectors are $(-0.7465, 0.6654)^T$ for
$c_{2}$ (non-KPZ mode) and $(0.6654, 0.7465)^T$ for $c_{1}$
(KPZ mode).

\begin{figure}[htbp]
\begin{center}
 {\includegraphics[width=0.45\textwidth]{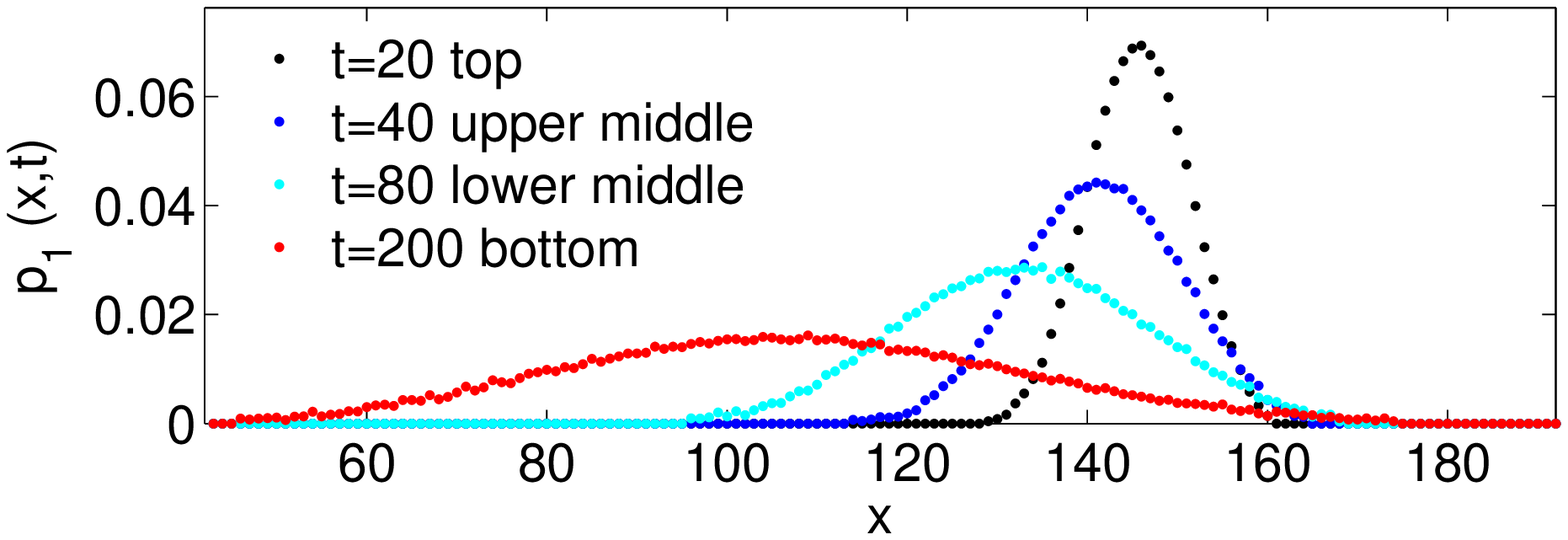}} 
{\includegraphics[width=0.45\textwidth]{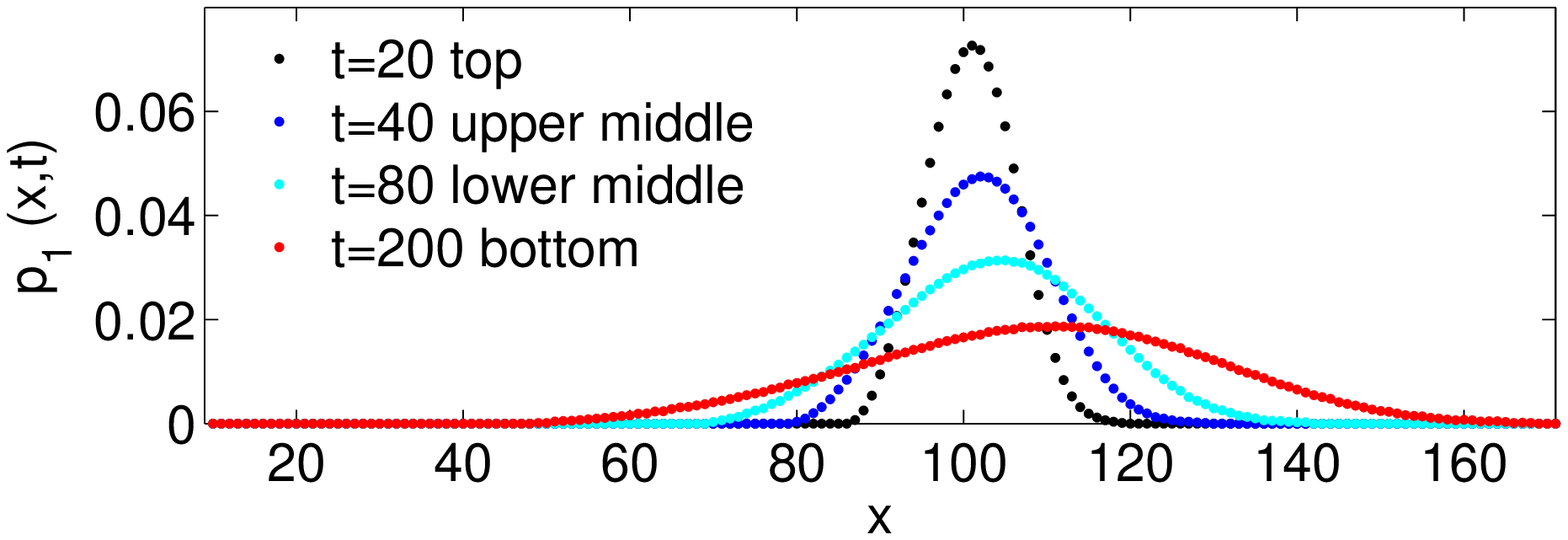}}
\caption{(Colour online) Case (ii): Dynamical structure functions for
particles on chain 1, for the KPZ mode (top) and the
non-KPZ mode (bottom) at different times from
Monte Carlo simulations, averaged over $10^7$($17 \cdot 10^7$)
histories for the KPZ (non-KPZ) mode for
$N=300$ ($N=200$).  Statistical errors are smaller than symbol size.}%
\label{FigKPZheatPeaks}
\end{center}
\end{figure}

\begin{figure}[htbp]
\begin{center}
 {\includegraphics[width=0.35\textwidth, height=35mm]{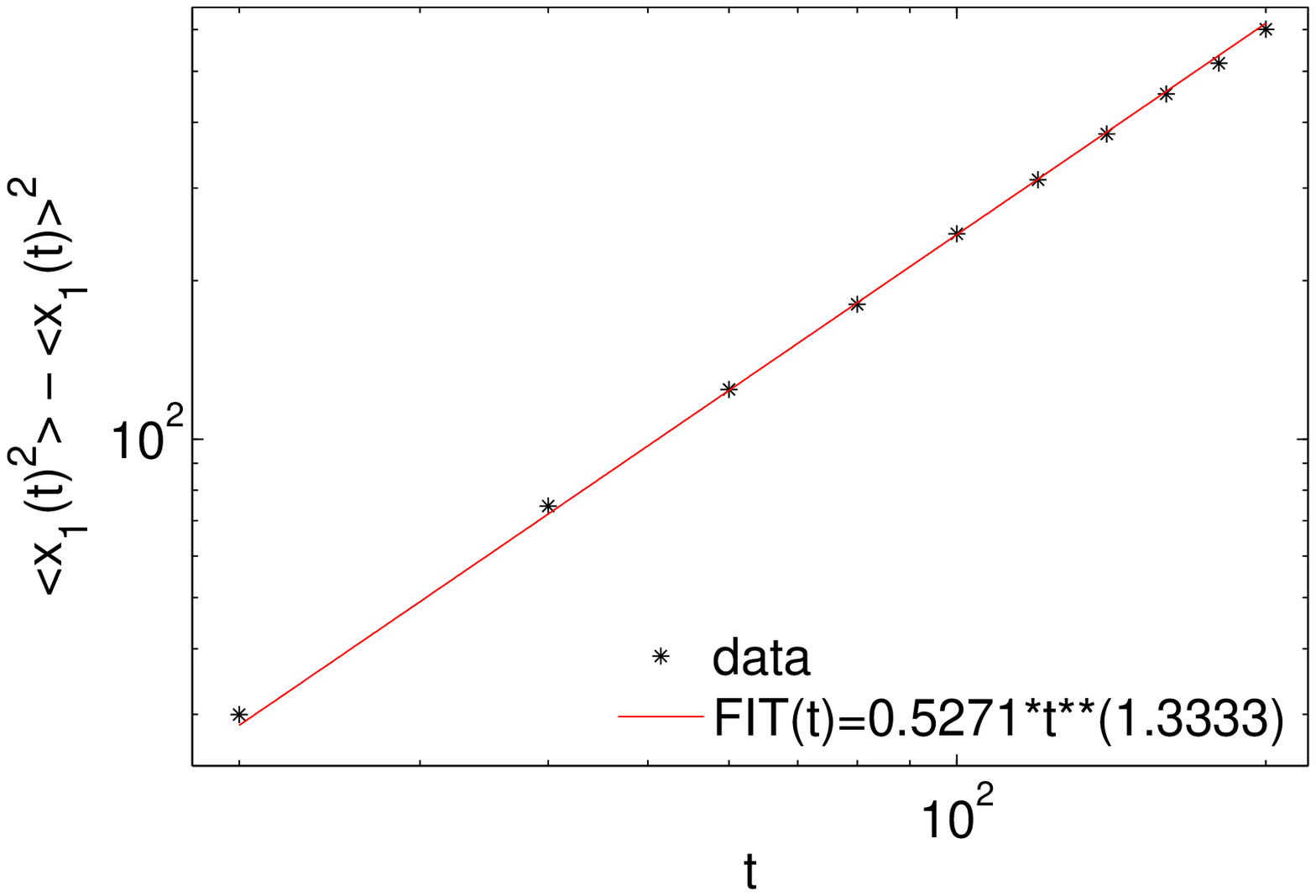}} 
{\includegraphics[width=0.35\textwidth, height=35mm]{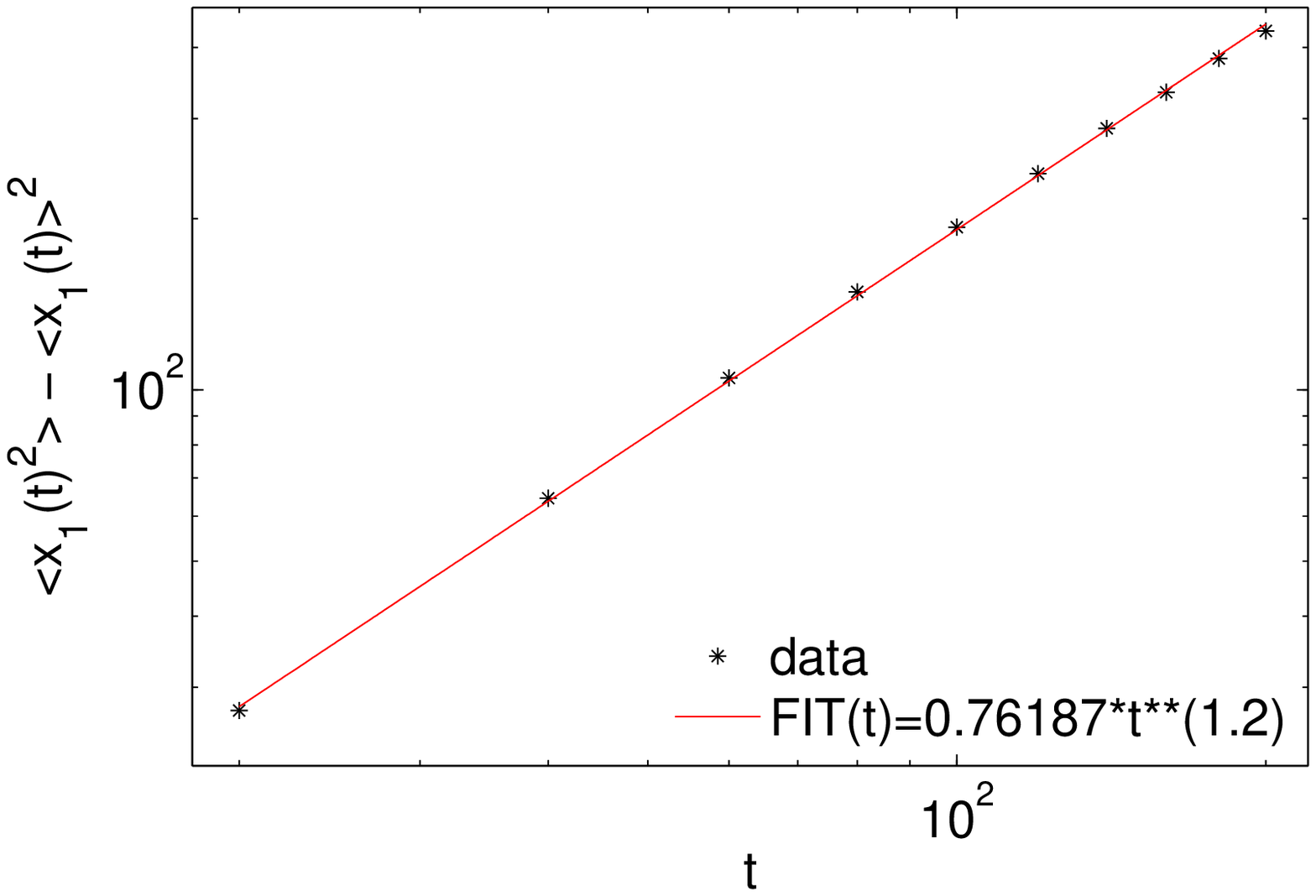}}
\caption{(Colour online) Case (ii). Variance of the dynamical structure function shown in Fig. 2. 
as function of time. The lines with the predicted universal
slopes $2/z=4/3$ for KPZ mode (top)  and $2/z=6/5$ for the non-KPZ
mode (bottom) are guides to the eye. Error bars (not shown) are approximately  symbol size.}%
\label{FigKPZheatSigma}
\end{center}
\end{figure}

The simulations confirm the predictions, see Figs. \ref{FigKPZheatPeaks} and \ref{FigKPZheatSigma}. 
For both modes the measured
velocity differs from the theoretical prediction by less than $0.003$. A linear least-square
fit on log-log scale of the simulation results for the variance of the non-KPZ mode 2
yields $2/z_2^{MC}=1.19 \pm 0.02$, very close to the mode-coupling value $2/z_2=6/5=1.2$. 
For the amplitude $\propto t^{-1/z}$ at the maximum as a function of time we find $1/z_2^{MC}=0.58$, 
also in good agreement with $1/z_2=0.6$. The fitted exponent 
$2/z_1^{MC}=1.302$ of the
KPZ mode 1 deviates slightly from $4/3$, which is consistent with
the strong coupling to the non-KPZ mode: The matrix element $G^{(1)}_{22}\approx 0.63$ is larger
than the KPZ self-coupling constant $G^{(1)}_{11}\approx 0.59$.


In order to test case (iii) (KPZ and diffusive mode), we choose
$\gamma=-0.8,b=1,\rho_{1}=\rho_{2}=0.5$. The characteristic velocities are
$c_1 = -0.2$ (eigenvector $(1,1)^T/\sqrt(2)$) and $c_1 = 0.2$ (eigenvector $(1,-1)^T/\sqrt(2)$).
The mode coupling matrices 
are given by
\[
G^{1}= 0.2121 \left(
\begin{array}
[c]{cc}%
1 & 0\\
0 & 1
\end{array}
\right), \quad
G^{2}=0.2121 \left(
\begin{array}
[c]{cc}%
0 & 1\\
1 & 0
\end{array}
\right)
\]
corresponding to a KPZ mode 1 and diffusive mode 2, see Figs. 4 and 5.
The characteristic velocities agree with the theoretical prediction
with an accuracy of better than $1 \%$ and also the measured scaling exponents 
$2/z_1^{MC}=1.343$, $2/z_2^{MC}=1.030$
are in good agreement with
the theoretical prediction $2/z_1=4/3$ and $2/z_2=1$.
We have also verified numerically the 
occurrence of two KPZ modes for generic values of the densities. This behaviour
is expected and data are not shown here.

\begin{figure}[htbp]
\begin{center}
{\includegraphics[width=0.45\textwidth]{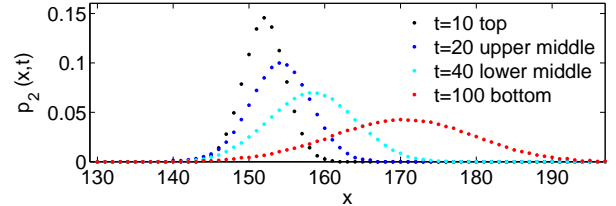}}
\caption{(Colour online) Case (iii): Dynamical structure function of the diffusive mode for
particles on chain 2 with
$N=300$, with $c_2=0.2$ (bottom) at different times $t$, from
Monte Carlo simulations, averaged over $10^7$
histories.  Statistical errors are smaller than symbol 
size.}%
\label{FigKPZdiffusionPeaks}
\end{center}
\end{figure}

\begin{figure}[htbp]
\begin{center}
{\includegraphics[width=0.35\textwidth, height=35mm]{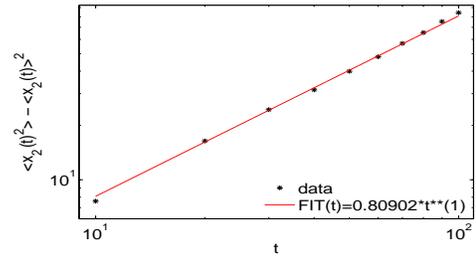}}
\caption{(Colour online) Case (iii). Variance of the dynamical structure function shown in Fig. 4. 
as function of time. The line with the predicted universal
slope $2/z=1$ for the diffusive
mode (bottom) are guides to the eye. Error bars (not shown) are approximately symbol size.}%
\label{FigKPZdiffusionSigma}
\end{center}
\end{figure}

In summary, we have shown that the two-lane asymmetric simple exclusion process with two conservation laws
exhibits anomalous transport and has a superdiffusive non-KPZ mode with dynamical exponent $z=5/3$ on a line in the space
of conserved densities $(\rho_1,\rho_2)$, provided that the model
is not symmetric under lane change. In the latter case of higher
symmetry a diffusive mode can occur instead of the non-KPZ mode. This is surprising as the hopping of 
both particle species is totally
asymmetric. We did not find
any point in parameter space where the KPZ mode would be completely absent.
We argue that the existence of a superdiffusive non-KPZ mode is generic for driven diffusive systems 
with more than one conservation law and will generally occur at some specific 
manifold in the space of conserved densities $\rho_i$.  This new universality class
for anomalous transport in driven diffusive systems is expected to result in a novel exponent
for the stationary density profile in open systems \cite{Krug91}.
An interesting open problem
that is raised by our findings is the role of symmetries for the suppression of the non-KPZ mode
and the occurence of a diffusive mode.

\begin{acknowledgments}
V.P. acknowledges financial support by DFG. 
G.M.S. thanks H. Spohn and H. van Beijeren for most illuminating discussions at TU Munich and
at the Oberwolfach workshop Large Scale Stochastic Dynamics. We also thank them and C. Mendl for useful 
comments on the
manuscript.
\end{acknowledgments}


\begin{thebibliography}{99}

\bibitem{Alde67}
B.J. Alder and T.E. Wainwright, 
Phys. Rev. Lett. {\bf 18}, 988 (1967).

\bibitem{Erns76}
M.H. Ernst, E.H. Hauge, and J.M.J. van Leeuwen, 
J. Stat. Phys. {\bf 15}, 7 (1976).

\bibitem {Prae02}
M. Pr\"ahofer and H. Spohn, in: \textit{In and Out of
Equilibrium}, edited by V. Sidoravicius, Vol. 51 of Progress in Probability
(Birkhauser, Boston, 2002).

\bibitem{Kard86}
M. Kardar, G. Parisi, and Y.-C. Zhang, 
Phys. Rev. Lett. {\bf  56}, 889 (1986).

\bibitem{Prae04}
M. Pr\"ahofer and H. Spohn,
J. Stat. Phys. {\bf 115}, 255 (2004).

\bibitem{Gras02}
P. Grassberger, W. Nadler, and L. Yang,
Phys. Rev. Lett. {\bf  89}, 180601 (2002).

\bibitem{Delf07}
L. Delfini, S. Lepri, R. Livi and A. Politi,
J. Stat. Mech. P02007 (2007).

\bibitem{Bern13}
C. Bernardin and P. Gon\c calves,
arXiv:1205.1879v3 (2013).

\bibitem{Spoh13}
C.B. Mendl and H. Spohn, Phys. Rev. Lett. {\bf 111}, 230601 (2013).

\bibitem{vanB12}
H. van Beijeren, 
Phys. Rev. Lett. {\bf 108}, 108601 (2012).

\bibitem{Ferr13}
P.L. Ferrari, T. Sasamoto and H. Spohn,
J. Stat. Phys.  \textbf{153}, 377--399 (2013).

\bibitem{Arit09}
C. Arita, A. Kuniba, K. Sakai and T. Sawabe,
J. Phys. A: Math. Theor. {\bf 42} 345002 (2009).

\bibitem{Das01}
D. Das,  A. Basu, M. Barma, and S. Ramaswamy, 
Phys. Rev. E \textbf{64}, 021402 (2001).

\bibitem{Rako04} 
A. R\'akos and G.M. Sch\"utz,
J. Stat. Phys. {\bf 117}, 55-76 (2004).


\bibitem{Popk03} V. Popkov and G.M. Sch\"utz, 
J. Stat. Phys. {\bf 112}, 523-540 (2003).

\bibitem{Popk04}
V. Popkov and M. Salerno, 
Phys. Rev. E \textbf{69}, 046103 (2004).

\bibitem{Kipn99}
C. Kipnis and C. Landim,
\newblock {\it Scaling limits of interacting particle systems} 
\newblock (Springer, Berlin, 1999)

\bibitem{Gris11}
R. Grisi and G.M. Sch\"utz,
J. Stat. Phys. \textbf{145}, 1499--1512 (2011).

\bibitem{Toth03}
B. T\'oth and B. Valk\'o,
J. Stat. Phys. \textbf{112}, 497--521 (2003).

\bibitem{Krug91}
J. Krug,
Phys. Rev. Lett. {\bf 67}  1882 (1991).


\end{thebibliography}
\end{document}